\documentstyle[12pt]{article}
\newcommand{\be}{\begin{equation}}
\newcommand{\ee}{\end{equation}}
\newcommand{\bea}{\begin{eqnarray}}
\newcommand{\eea}{\end{eqnarray}}
\newcommand{\ba}{\begin{array}}
\newcommand{\ea}{\end{array}}

\def\bbox{{\,\lower0.9pt\vbox{\hrule \hbox{\vrule height 0.2 cm
\hskip 0.2 cm \vrule height 0.2 cm}\hrule}\,}}
\newcommand{\dsl}{\pa \kern-0.5em /}

\newcommand{\subskip}{\vskip 0.1in\noindent}

\begin{document}

%%%%%%%%%%%%%%%% title page %%%%%%%%%%%%%%%%%%%%%%%%%%%%%%%%%%%%

\begin{titlepage}
\vfill
\begin{flushright}
hep-th/0009062\\
\end{flushright}

\vfill

\begin{center}
\baselineskip=16pt
{\Large\bf Calibrations and Fayyazuddin-Smith Spacetimes}
\vskip 1.0cm
%{\Large\bf Fayyazuddin-Smith Spacetimes}
\vskip 0.3cm
{\large {\sl }}
\vskip 10.mm
{\bf Hyunji Cho, Moataz Emam, David Kastor and Jennie Traschen} \\
%\\[2mm]
\vskip 1cm
%\vfill
{

  Department of Physics\\
  University of Massachusetts\\
  Amherst, MA 01003\\
}
\vspace{6pt}
\end{center}
\vskip 1.0in
\par
\begin{center}
{\bf ABSTRACT}
\end{center}
\begin{quote}
We show that a class of spacetimes introduced by Fayyazuddin and Smith to describe intersecting M5-branes admit a generalized Kahler calibration.  Equipped with this understanding, we are able to construct spacetimes corresponding to further classes of calibrated $p$-brane world-volume solitons.  We note that these classes of spacetimes also describe the fields of $p$-branes wrapping certain supersymmetric cycles of Calabi-Yau manifolds.
\vfill
% \hrule width 5.cm
\vskip 2.mm
\end{quote}
\end{titlepage}
%%%%%%%%%%%%%%%%%%%%%%%%%%%%%%%%%%%%%%%%
\setcounter{equation}{0}

\section{Introduction}

The mathematical theory of calibrations \cite{Harvey:1982} and certain extensions thereof have proven to be quite useful in classifying the supersymmetric worldvolume solitons \cite{Gibbons:1998xz,Callan:1998kz} of branes embedded in fixed supersymmetric spacetime backgrounds.  Applications include smoothed intersections of branes in flat spacetime \cite{Gibbons:1999hm,Gauntlett:1999vk,Acharya:1998en,Acharya:1998st,Figueroa-O'Farrill:2000su,Papadopoulos:2000yx}; branes wrapping supersymmetric cycles of Calabi-Yau manifolds \cite{Becker:1995kb,Becker:1996ay}; worldvolume solitons in AdS compactifications \cite{Gutowski:1999iu}; solitons involving worldvolume gauge fields that describe branes ending on other branes \cite{Gauntlett:1999aw,Barwald:1999ux,Gauntlett:1999wb}; and branes in $p$-brane spacetime backgrounds \cite{Gutowski:1999tu}.
  
The calibrated branes in each of these applications have so far been treated as test objects.  In this paper we demonstrate that the calibration technology is also useful in understanding the spacetime fields that result from treating these worldvolume solitons as charged, gravitating sources.  The reason for this is quite simple. 
Consider the spacetime geometry generated by a supersymmetric worldvolume soliton.  
Based on the BPS `no force' properties of branes, it should be possible for a suitably configured test brane embedded in this spacetime to be in equilibrium.  The spacetime should therefore carry a calibrating form.  Moreover, near infinity,  this calibrating form should approach the fixed background form that calibrated the original worldvolume soliton.  This situation holds, for example, for the spacetime of a single planar $M$-brane \cite{Gutowski:1999tu}\footnote{It was also demonstrated in \cite{Gauntlett:1999xz} that the BPS spike soliton of a test D$3$-brane, describing a fundamental string ending on the brane, can also be found when the test D$3$-brane is placed in a D$3$-brane spacetime background}.

Our starting point will be a class of supersymmetric spacetimes constructed by Fayyazuddin and Smith to describe the spacetime fields of M$5$-branes intersecting on $3$-branes \cite{Fayyazuddin:1999zu}.  A central ingredient in these spacetimes is a warped Kahler metric living on the four relative tranverse directions of the intersecting brane configuration.  The Kahler metric depends, as well, on the overall transverse coordinates.  The exact form of the Kahler metric and the warp factor are related to the M$5$-brane sources by a nonlinear field equation.  We will see that these spacetimes indeed have calibrating forms of the appropriate type and that, equipped with this understanding, one can construct similar spacetimes corresponding to other calibrated worldvolume solitons.  We will refer to these different spacetimes collectively as Fayyazuddin-Smith (FS) spacetimes.  We will focus here on Kahler calibrations and correspondingly on FS spacetimes built around Kahler metrics.  More generally, FS spacetimes will involve other metrics of reduced holonomy.  We conjecture that the spacetime fields of all calibrated worldvolume solitons will be of FS type.

Of course, it has proven to be quite difficult to construct spacetimes corresponding to particular configurations of localized intersections of branes \cite{Tseytlin:1997cs,Itzhaki:1998uz,Surya:1998dx,Fayyazuddin:1999zu,Youm:1999zs,Gomberoff:2000ps,Cvetic:2000cj}, as opposed to smeared intersections (see \cite{Youm:1999hw} for a complete review).  It was argued in \cite{Marolf:1999uq} that this situation may reflect interesting underlying physics.  The world-volume effective field theory description of the delocalization of certain brane intersections is related to the Coleman-Mermin-Wagner theorem.  In these cases, the dimensionality of the intersection is the determining factor as to whether the localisation of the classical world-volume soliton persists in the supergravity solution.  FS spacetimes should provide the appropriate supergravity setting to study these effects.

We note that FS spacetimes also provide the spacetime fields of branes wrapping supersymmetric cycles of Calabi-Yau manifolds.   To describe intersecting branes in otherwise empty spacetime, the $4$ real dimensional Kahler metric in the original FS ansatz \cite{Fayyazuddin:1999zu} is taken to be asymptotic to $4$-dimensional flat space.  However, if instead it is taken to be asymptotic to a Calabi-Yau metric, {\it e.g.} to a Ricci flat metric on K3, then the FS spacetimes \cite{Fayyazuddin:1999zu} describe M5-branes wrapping supersymmetric (1,1) cycles of K3\footnote{See \cite{Maldacena:2000mw} for a recent related discussion of branes wrapping cycles of K3.}.  The spacetime geometry of branes wrapping all of K3 has been shown to reflect very interesting underlying physics \cite{Johnson:2000qt}.  It seems likely that FS spacetimes will provide a rich ground for further study in this context.  If we take, for example, M$2$-branes wrapping $2$-cycles of compact Calabi-Yau $3$-folds, then from the $5$-dimensional viewpoint these will be black holes.  Dimensionally reducing and keeping only the massless Kaluza-Klein modes should give the black holes of \cite{Chou:1997ba,Sabra:1998yd,Chamseddine:1998yv,Gaida:1999km,Chamseddine:1999qs}.  The FS spacetimes should provide the $11$-dimensional lifts of these spacetimes, including the nontrivial massive Kaluza-Klein modes as well.

Finally, we note that another extension of the FS class of spacetimes to include branes ending on branes has recently been given in \cite{Rajaraman:2000ws}.  We expect that these spacetimes may also be usefully organized using calibration technology.

\section{Calibrations}

We start with a brief and basic introduction to calibrations.
Consider the action for a $p$-brane moving in a $D+1$ dimensional spacetime with metric $G_{\mu\nu}$ and $(p+1)$-form gauge potential $A_{\mu_1\dots\mu_{p+1}}$,
\be\label{action}
S_{p+1}=\int d^{p+1}\sigma\left\{
\sqrt{-\det g_{ab}}- {1\over (p+1)!}\varepsilon^{a_1\dots a_{p+1}}
\partial_{a_1}X^{\mu_1}\dots\partial_{a_{p+1}}X^{\mu_{p+1}}
A_{\mu_1\dots\mu_{p+1}}\right\},
\ee
where $\sigma^a$ with $a=0,1,\dots,p$ are world-volume coordinates, $X^\mu(\sigma)$ gives the embedding of the brane in the background spacetime and $g_{ab}=\partial_aX^\mu\partial_bX^\nu G_{\mu\nu}$ is the induced metric on the world-volume.  We will not consider here possible world-volume gauge fields, or couplings to additional spacetime fields. To start, let us assume a flat background, $G_{\mu\nu}=\eta_{\mu\nu}$, $A_{\mu_1\dots\mu_{p+1}}=0$ and consider static brane configurations.  These will minimize the spatial volume of the brane.  Calibrations are a mathematical technique for finding classes of such minimal submanifolds.  A calibration for a $p$-dimensional submanifold is a $p$-form $\phi$ on the embedding space that satisfies two properties

\begin{enumerate}
\item The calibration $\phi$ is a closed form
\be\label{closed}
d\phi=0.
\ee

\item The pullback of $\phi$ onto any $p$-dimensional submanifold $\Sigma$ is always less than, or equal to, the induced volume form on the submanifold,
\be\label{inequality}
{}^*\phi\le \varepsilon_\Sigma.
\ee
\end{enumerate}
It then follows via a simple argument that, if the inequality (\ref{inequality}) is saturated at every point on a $p$-dimensional submanifold $\Sigma$, then $\Sigma$ minimizes volume within its homology class.  Assume $\Sigma$ saturates the inequality (\ref{inequality}) at every point.  Pick a closed $(p-1)$-dimensional surface $S$ in $\Sigma$, and within $S$ continuously deform $\Sigma$ into a new submanifold $\Sigma^\prime$.  The following chain of equalities and inequalities then shows that $Vol(\Sigma)\le Vol(\Sigma^\prime)$,
\be\label{chain}
Vol(\Sigma) = \int_\Sigma\varepsilon_\Sigma   = \int_\Sigma {}^*\phi 
= \int_{\Sigma^\prime} {}^*\phi +\int_B d\phi  = \int_{\Sigma^\prime} {}^*\phi
\le \int_{\Sigma^\prime}\varepsilon_{\Sigma^\prime} = Vol(\Sigma^\prime),
\ee
where $B$ is the $p$-dimensional region bounded by $\Sigma$ and $\Sigma^\prime$.

\vskip 0.1in
\noindent{\bf Kahler Calibrations:} The simplest examples of calibrating forms and the ones that will concern us below are the Kahler calibrations.  Start with even dimensional flat space, $D=2n$, with real Cartesian coordinates $x^1,\dots,x^{2n}$.  Choose a complex structure, {\it i.e.} a pairing of real coordinates into complex coordinates, for example,
\be
z^1=x^1+ix^2,\dots, z^n=x^{2n-1}+ix^{2n},
\ee
then the Kahler form is given by
\be
\ba{rl}
\omega=&dx_1\wedge dx_2+\dots +dx_{2n-1}\wedge dx_{2n}\\
=&{i\over 2}\left(dz_1\wedge d\bar z_1+\dots dz_n\wedge d\bar z_n\right).\\
\ea
\ee
The forms $\phi_{2k}=\omega^k/k$ can then be shown to be calibrations \cite{Harvey:1982}.  The corresponding calibrated submanifolds are simply the complex submanifolds of real dimension $2k$.  

We recall some examples of calibrated surfaces \cite{Gibbons:1999hm,Gauntlett:1999vk} that will be useful to keep in mind below.  Our focus below will be with M2-branes and M5-branes and we frame the examples in this context. First consider a static M2-brane configuration whose world-volume lies entirely in the $(1,2,3,4)$ subspace of $10$ dimensional flat space. 
We can then take $D=4$ above and the calibrating $2$-form $\phi=\omega=dx_1\wedge dx_2 +dx_3\wedge dx_4$.  Clearly if the M2-brane to lies either in the $(1,2)$ plane, or in the $(3,4)$ plane, then the inequality (\ref{inequality}) is saturated and these are calibrated surface.  A more nontrivial example is the family of complex curves
\be\label{smooth}
z_1z_2=\alpha^2,
\ee
with $\alpha$ an arbitrary constant.  These curves interpolate smoothly between the $(1,2)$ and $(3,4)$ planes and represent a smoothed version of two static M$2$-branes intersecting at a point.  The singular limit $\alpha=0$ gives the pure orthogonal intersection of the two planes.  If we added on $3$ additional flat spatial directions to the brane, then the curve (\ref{smooth}) gives two $M5$-branes intersecting on a $3$-brane.

Now take $D=6$ and consider the $4$-form calibration $\phi={1\over 2}\omega\wedge\omega$.  In terms of the real coordinates this is
\be
\phi=dx_1\wedge dx_2\wedge dx_3\wedge dx_4 +
dx_1\wedge dx_2\wedge dx_5\wedge dx_6 +
dx_3\wedge dx_4\wedge dx_5\wedge dx_6.
\ee
Clearly the $(1,2,3,4)$, $(1,2,5,6)$ and $(3,4,5,6)$ planes are examples of submanifolds calibrated by $\phi$, and complex surfaces exist that interpolate smoothly between these planes. Adding another spatial direction $x^7$ to get $M5$-branes, there will then be calibrated surfaces describing the smoothed intersection of three $5$-branes in the directions
\be\label{M5-branes}
\ba{l}
(t,1,2,3,4,x,x,7)\\
(t,1,2,x,x,5,6,7)\\
(t,x,x,3,4,5,6,7)\\
\ea
\ee
where the `x's are placeholders.  Note that each pair of M$5$-branes intersects on a $3$-brane and that altogether they intersect on a string.

\subskip{\bf Calibrations \& Spinors:}
The calibration technology applies in curved spaces as well.  For example, if $\omega$ is now the Kahler form for an arbitrary Kahler space, then the forms $\phi_{2k}=\omega^k/k$ are again calibrations and the calibrated submanifolds are again the set of complex submanifolds.  In general the existence of calibrating forms is tied to the property of reduced holonomy (see {\it e.g.} \cite{bryantnotes}).  Reduced holonomy in turn is tied to the existence of spinor fields having special properties. 

For example, for an $N$ complex dimensional Kahler manifold with metric $g_{m\bar n}$, the holonomy group is $U(N)\subset SO(2N)$.  Covariantly constant spinors exist only in the Calabi-Yau case of vanishing Ricci tensor, for which the holonomy group is further reduced to $SU(N)$. For a general Kahler metric, though, there exists a pair of spinors $\epsilon_+$ and $\epsilon_-$ transforming as singlets of the holonomy group.  These satisfy the relations 
\be\label{annihilate}
\Gamma_m\epsilon_+=\Gamma_{\bar m}\epsilon_-=0,
\ee
from which follow the projection conditions
\be\label{kahlerproj}
\Gamma_{m\bar n}\epsilon_\pm=\pm g_{m\bar n}\epsilon_\pm .
\ee
If we normalize $\epsilon_\pm^\dagger\epsilon_\pm=1$ then the Kahler form can be written as 
\be\label{kahlerform}
\omega_{ab}=\pm i\epsilon_\pm^\dagger\Gamma_{ab}\epsilon_\pm.
\ee
The vanishing of the components $\omega_{mn}$ and $\omega_{\bar m\bar n}$ follow from the relations (\ref{annihilate}), which also imply that only even dimensional forms with equal numbers of holomorphic and anti-holomorphic indices can be built in this way.  The covariant derivatives of $\epsilon_\pm$ are given by
\be\label{kahlerderiv}
\nabla_p\epsilon_\pm=\partial_p\epsilon_\pm \pm {1\over 2}(\bar E^{-1}\partial_p\bar E)\epsilon_\pm,\qquad
\nabla_{\bar p}\epsilon_\pm=\partial_{\bar p}\epsilon_\pm \mp {1\over 2}(E^{-1}\partial_{\bar p} E)\epsilon_\pm ,
\ee
where $E$ and $\bar E$ are determinants of the complex frame fields $E^{\hat m}_n$ and $E^{\hat{\bar m}}_{\bar n}$ respectively, with hat denoting flat space frame indices.  For the Ricci flat case, these second terms vanish identically giving covariantly constant spinors.

\vskip 0.1in
\noindent{\bf Generalized Calibrations:} A certain amount of care is neccesary in applying the calibration technology to find static solutions for $p$-branes in curved spacetimes \cite{Gutowski:1999tu}, because even for a static $p$-brane, the time-time component $G_{00}$ of the spacetime metric enters the $p$-brane effective action (\ref{action}).  Assume that the embedding spacetime is static with timelike Killing vector $\xi^a=({\partial\over \partial x^0})^a$.  If we fix static gauge $\sigma^0=x^0$ for the coordinates on the brane worldvolume, then the $\sqrt{-\det g_{ab}}$ term in the brane action (\ref{action}) includes a contribution from $g_{00}=G_{00}$, called the redshift factor in \cite{Gutowski:1999tu}.  This factor can be absorbed by defining a new effective spatial metric $\hat G_{\alpha\beta}= (-G_{00})^{1/p}G_{\alpha\beta}$ where $\alpha,\beta=1,\dots,D$ now run over only spatial directions in the embedding space.  We then have
\be
\sqrt{-\det g_{ab}}=\sqrt{\det\hat g_{kl}}
\ee
where $k,l=1,\dots p$ are purely spatial world-volume indices and $\hat g_{kl}$ is the spatial metric induced on the brane via embedding in the rescaled metric $\hat G_{\alpha\beta}$ defined above.  If there are additional spatial symmetry directions of the embedding space that are shared by the $p$-brane configuration, then these can be handled in a similar manner \cite{Gutowski:1999tu} by appropriately modifying the definitions of $\hat G_{\alpha\beta}$, $\hat A$ and $\hat F$.

Finally, if the spacetime has a nonzero $(p+1)$-form gauge potential $A_{\mu_1\dots\mu_{p+1}}$, then a static brane configuration will satisfy equations of motion involving the corresponding field strength.  An appropriately generalized definition of calibrating forms taking this additional force into account was given in \cite{Gutowski:1999iu,Gutowski:1999tu}.  The modification required is quite simple.
Condition (\ref{closed}) becomes 
\be\label{generalized}
d\phi = \hat F
\ee
where $\hat F=d \hat A$ and  
$\hat A_{\alpha_1\dots\alpha_p}=A_{0\alpha_1\dots\alpha_p}$.  Therefore the calibrating form $\phi$ is equal to the reduced gauge potential $\hat A$ up to a gauge transformation.  This new condition then yields a chain of equalities and inequalities similar to (\ref{chain}), showing that if a static surface saturates the calibration bound then it minimizes the action (\ref{action}).  

\vskip 0.1in
\noindent{\bf M$2$-Brane Spacetime:} The planar M$2$-brane itself provides a good example of a spacetime with a generalized calibrating form \cite{Gutowski:1999tu}.
\be\label{m2brane}
ds^2=H^{-2/3}(-dt^2+dx_1^2+dx_2^2)+H^{+1/3}(dx_3^2+\dots +dx_{10}^2),\qquad A_{t12}=cH^{-1},
\ee
where $c=\pm 1$.  For a static test M2-brane in this background the effective spatial metric and gauge potential defined above are given by
\be\label{rescaled}
d\hat s^2=H^{-1}(dx_1^2+dx_2^2)+dx_3^2+\dots +dx_{10}^2 ,\qquad \hat A_{12}=cH^{-1}.
\ee
As discussed in \cite{Gutowski:1999tu}, test M2-branes will then be calibrated by the form
\be\label{calib}
\phi=cH^{-1}dx_1\wedge dx_2+\omega_\perp,
\ee
where $\omega_\perp$ is an arbitrary Kahler form on the transverse space, equivalent to a choice of complex structure in the transverse space.  The calibrating form $\phi$ is then gauge equivalent to the gauge potential $\hat A$ and equation (\ref{generalized}) is satisfied.  The calibrated surfaces are complex surfaces with respect to the associated almost complex structure obtained by raising one index on $\phi_{kl}$ using the rescaled metric (\ref{rescaled}).  Note that the warp factor $H$ then drops out.

\section{Fayyazuddin-Smith Spacetimes}

The original FS spacetimes \cite{Fayyazuddin:1999zu} described M5-branes intersecting on 3-branes.  Here we start with the related M$2$-brane FS spacetimes studied in \cite{Gomberoff:2000ps}. The metric and gauge potential for these are given by
\be\label{original}
\ba{l}
ds^2=H^{-2/3}\left(-dt^2+2g_{m\bar n}dz^mdz^{\bar n}\right) + H^{+1/3}(\delta_{\alpha\beta}dx^\alpha dx^\beta)\\
A_{tm\bar n}= icH^{-1}g_{m\bar n},\qquad c=\pm 1\\
\ea
\ee
Here $z^m$ with $m=1,2$ are complex coordinates on a 4 real dimensional Kahler manifold ${\cal M}$ and $\alpha,\beta=5,\dots,10$ are indices for the $6$-dimensional transverse space.  The Kahler metric $g_{m\bar n}$ on ${\cal M}$ is allowed to depend on the transverse coordinates $x^\alpha$ as well as on position in ${\cal M}$, {\it i.e.} it can be written as $g_{m\bar n}=\partial_m\partial_{\bar n}K(z^p,\bar z^q, x^\alpha)$  with $K$ a Kahler potential depending on the transverse coordinates.  The warp factor $H$ is also allowed to depend on both position in ${\cal M}$ and position in the transverse space.  Note that the Kahler metric $g_{m\bar n}$ at a fixed transverse position is not required to be Ricci flat.

We present a  detailed review of the supersymmetry conditions in order to correct a mistake in the form of the results stated in \cite{Fayyazuddin:1999zu,Gomberoff:2000ps} that has obstructed a better understanding of this class of spacetimes.  
The supersymmetry condition for $D=11$ supergravity takes the form $\hat\nabla_A\epsilon=0$, where 
\be\label{supercovariant}
\hat\nabla_A\epsilon=\nabla_A\epsilon +
{1\over 288}\left(\Gamma_A^{\ BCDE}-8\delta_A^B\Gamma^{CDE}\right)F_{BCDE}\epsilon ,
\ee
and $(A,B,\dots)$ are $D=11$ indices.
The supercovariantly constant spinors of the FS spacetimes (\ref{original}) satisfy the projection conditions
\be\label{projection}
\ba{c}
\Gamma^{\hat m\hat{\bar n}}\epsilon= a\delta^{\hat m\hat{\bar n}}\epsilon, \qquad a=\pm 1 ,\\
\Gamma^{\hat t}\epsilon= ib\epsilon, \qquad b=\pm 1\\
\ea
\ee
where hatted indices are frame indices and the signs of the two projections are correlated with the sign of the gauge potential by the relation $abc=-1$.  The first projection condition is just the standard projection (\ref{kahlerproj}) onto singlets of the $U(2)$ holonomy group of the Kahler metric $g_{m\bar n}$ at fixed transverse position.  The combination of the two projections reduces the fraction of supersymmetry preserved to $1/4$.

Given the projections (\ref{projection}), the supersymmetry conditions then impose a relation between the warp factor $H$ and the complex determinant of the Kahler metric 
$g=g_{1\bar 1}g_{2\bar 2}-g_{1\bar 2}g_{2\bar 1}$.  
In \cite{Gomberoff:2000ps}, and originally in \cite{Fayyazuddin:1999zu} for the M$5$-brane case, this relation is given as $H=g$.  However, this is not precisely correct, as the following argument shows. 
The form of the FS ansatz (\ref{original}) is preserved by holomorphic coordinate transformations on the Kahler manifold that do not depend on the transverse coordinates, 
\be\label{transformations}
z^{\prime m}=z^{\prime m}(z^p).
\ee
Under these transformations the warp factor $H$ is invariant, but the determinant $g$ is transformed to $g^\prime=gf\bar f$ where $f$ is holomorphic.  Hence the condition $H=g$ is not covariant under the transformations (\ref{transformations}).

In order to determine the correct conditions, we write out the requirements that each of the components of $\hat\nabla_A\epsilon=0$ reduce to after having applied the projection conditions (\ref{projection})
\be\label{derivs}
\ba{lc}
A=t: & \partial_\alpha\log H -\partial_\alpha\log g=0 \\
A=p: & \partial_p\epsilon+\left\{\left({1\over 6}
+{a\over 4}\right)\partial_p\log H -{a\over 2}\partial_p\log\bar E \right\}\epsilon=0 \\
A=\bar p: & \partial_{\bar p}\epsilon+\left\{\left({1\over 6}
-{a\over 4}\right)\partial_{\bar p}\log H +{a\over 2}\partial_{\bar p}\log E \right\}\epsilon =0 \\
A=\alpha: & \partial_\alpha\epsilon +\left\{{1\over 6}\partial_\alpha\log H +
{a\over 4}\partial_\alpha\log E
-{a\over 4}\partial_\alpha\log\bar E\right\}\epsilon=0 .\\
\ea
\ee
One can check that the whole set of equations can be solved provided that
\be\label{susy}
\ba{c}
\partial_\alpha\log H= \partial_\alpha\log g\\
\partial_m\partial_{\bar n}\log H=\partial_m\partial_{\bar n}\log g.\\
\ea
\ee
It is worth noting that the second condition involves the Kahler metric at fixed transverse position through its Ricci tensor ${\cal R}_{m\bar n}=-\partial_m\partial_{\bar n}g$, which transforms as a tensor under the coordinate transformations (\ref{transformations}).  These new relations are then covariant under the holomorphic coordinate transformations (\ref{transformations}).  These conditions imply that the general form of the relation between $H$ and $g$ is given by
\be\label{realrelation}
g=Hf\bar f, 
\ee
where $f(z^m)$ is a holomorphic function of the complex coordinates and is independent of the transverse coordinates $x^\alpha$.  We then find using equation (\ref{realrelation}) that the supercovariantly constant spinors are then given by
\be
\epsilon=(E/f)^{-({1\over 6}+{a\over 4})}(\bar E/\bar f)^{-({1\over 6}-{a\over 4})}\epsilon_0,
\ee
where $\epsilon_0$ is a constant spinor satisfying the projections (\ref{projection}).  The supercovariantly constant spinors are invariant under the coordinate transformations (\ref{transformations}).

All spacetimes of the form (\ref{original}) satisfying the relations (\ref{susy}) are supersymmetric.  However, we have not yet imposed the gauge field equations of motion.  For the FS spacetimes (\ref{original}), these reduce to the equations
\be\label{gaugeeom}
2\partial_m\partial_{\bar n} H+\delta^{\alpha\beta}\partial_\alpha\partial_\beta g_{m\bar n}=0,
\ee
which are covariant with respect to the holomorphic coordinate transformations (\ref{transformations})\footnote{$D=11$ supergravity can be coupled to M-brane sources by combining the bulk supergravity action with the M2-brane and M5-brane Born-Infeld actions.  For M2-brane sources, the resulting current contribution to the right hand side of (\ref{gaugeeom}) is given in \cite{Gomberoff:2000ps}.}.  
Combining the gauge field equation of motion (\ref{gaugeeom}) with the condition (\ref{realrelation}) gives a set of coupled nonlinear equations that has proved difficult to solve.
Solutions have been given in the M5-brane case in the near horizon limit \cite{Fayyazuddin:1999zu,Fayyazuddin:2000em,newfs} and to first order in the far field limit \cite{Gomberoff:2000ps}.

The gauge field equation of motion (\ref{gaugeeom}) can be rewritten as an equation for the Ricci tensor of the Kahler metric ${\cal R}_{m\bar n}$ at fixed transverse position, giving
\be\label{ricci}
{\cal R}_{m\bar n}=
{(\partial_m H)(\partial_{\bar n}H)\over H^2}+{1\over 2H}\delta^{\alpha\beta}\partial_\alpha\partial_\beta g_{m\bar n}.
\ee
It is worth noting that given the correct relation (\ref{realrelation}) between $g$ and $H$, the standard supersymmetric supergravity vacua now solve the field equations.  If $g_{m\bar n}$ is Ricci flat and independent of the transverse coordinates, then one can choose complex coordinates so that $g=1$ everywhere.  Taking $H=g=1$ then clearly gives a solution of equation (\ref{ricci}).  Performing a holomorphic coordinate transformation as in (\ref{transformations}) on this spacetime yields $g=f\bar f$, with $f$ holomorphic, and $H=1$, which is still obviously a solution to (\ref{ricci}).  However, if we instead also change $H$, so that as in \cite{Fayyazuddin:1999zu} $H=g=f\bar f$, then the spacetime no longer solves equation (\ref{ricci}).  In this case, by referring to  equation (\ref{original}), we see that the gauge potential has nonzero field strength.
These spacetimes correspond to nontrivial configurations of M$2$-brane sources. This is consistent because these spacetimes are not related by coordinate transformations to the the original Ricci flat vacuum spacetime.

\section{Calibrations \& New M-Brane Spacetimes} 
We now want to look at the FS spacetimes from the point of view of calibrations.  The perspective we gain will prove useful in  finding FS spacetimes for other types of M-brane world-volume solitons.
It is straightforward to check that the FS spacetimes (\ref{original}) discussed above have generalized calibrating forms in the sense defined in \cite{Gutowski:1999iu,Gutowski:1999tu}.  The effective spatial metric and gauge potential seen by a static M$2$-brane probe are
\be\label{effective}
\ba{l}
d\hat s^2=2H^{-1}g_{m\bar n}dz^mdz^{\bar n} + 
\delta_{\alpha\beta}dx^\alpha dx^\beta\\
\hat A_{m\bar n}=ic H^{-1}g_{m\bar n},\qquad c=\pm 1\\
\ea
\ee
The corresponding generalized calibrating $2$-form is given by
\be\label{warpedform}
\phi=cH^{-1}\omega_{\cal M}+\omega_\perp,
\ee
where $\omega_{\cal M}=ig_{m\bar n}dz^m\wedge dz^{\bar n}$ is the Kahler form associated with the metric $g_{m\bar n}$, and $\omega_\perp$ is an arbitrary Kahler form on the transverse space.  The calibrated surfaces are complex surfaces with respect to the almost complex structure obtained by raising one index on $\phi$.  Note that the warp factor $H$ again drops out from the almost complex structure.

What can we learn from this structure that will be useful in constructing FS spacetimes for other types of world-volume solitons?  The FS spacetimes (\ref{original}) arise in two diferent physical contexts.  In \cite{Gomberoff:2000ps}, the FS spacetimes were considered to be generated by static M2-brane sources lying on a nontrivial holomorphic curves in a $4$ dimensional subspace of $D=10$ flat space.  The Kahler metric $g_{m\bar n}$ was taken to be flat near infinity in the transverse space.  A second application is to take the Kahler manifold ${\cal M}$ to be K3 and letting the Kahler metric $g_{m\bar n}$ approach a Ricci flat K3 metric near infinity.  The FS spacetimes then describe M2-branes wrapping $(1,1)$ cycles of K3.  In each of these cases the original source branes were calibrated by the corresponding Kahler forms of these supersymmetric vacua.  The warped Kahler form $\phi$ in (\ref{warpedform}) approaches the corresponding vacuum Kahler form near infinity, since as we have argued above, $H$ must approach unity near infinity.  

We conjecture that a similar structure will hold for spacetimes corresponding to other calibrated worldvolume solitons.  For Kahler calibrated solitons, we expect to find an FS spacetime built around a Kahler metric, with a gauge potential simply related to the original calibrating form.  For another type of calibrated worldvolume soliton, we would expect to find an FS spacetime built around a general curved space that admits this type of calibration.  For special Lagrangian solitons, for example, we would expect an FS spacetime built around a Ricci flat Kahler metric. For a soliton calibrated by an exceptional calibration, we expect to find an FS spacetime built by a warped construction around a space with the corresponding reduced holonomy.  Below, we give results for Kahler calibrated solitons.  We will return to the other cases in future work.  

\subskip{\bf New M2-Brane Spacetimes:} 
The most straightforward generalization of the FS construction is to increase the number of dimensions of the Kahler manifold, maintaining the same basic form of the FS spacetimes (\ref{original}).   For {\it e.g.} a $3$ complex dimensional space this would correspond to $1/8$ supersymmetric, smoothed intersections of $3$ M2-branes, or to M2-branes wrapping $(1,1)$ cycles of Calabi-Yau $3$-folds.  Setting the complex dimension to be $N$, we make the ansatz
\be\label{kahler}
\ba{l}
ds^2=-H^{-2A}dt^2+2H^{-2B}g_{m\bar n}dz^mdz^{\bar n} + H^{2C}(\delta_{\alpha\beta}dx^\alpha dx^\beta)\\
A_{tm\bar n}=ic H^{-1}g_{m\bar n},\qquad c=\pm 1\\
\ea
\ee
where now the complex coordinates $m,n=1,\dots,N$ and the transverse coordinates $\alpha,\beta=1,\dots,10-2N$.  The nontrivial possibilities are $N=2,3,4,5$.   These spacetimes describe either $1/2^N$ supersymmetric smoothed intersections of M2-branes in otherwise empty spacetime, or to M2-branes wrapping $(1,1)$ cycles of Calabi-Yau N-folds.  

We find that these spacetimes preserve $1/2^N$ supersymmetry, if the exponents are given by
\be\label{exponents}
A={1\over 3}(N-1),\qquad B={1\over 6}(4-N), \qquad C={1\over 6}(N-1)
\ee
and $H$, $g_{m\bar n}$ related in general as in equation (\ref{realrelation})\footnote{Note that for $N=1$ the exponents in (\ref{exponents}) yield flat Minkowski spacetime.  This seems puzzling because the ansatz (\ref{kahler}) should cover the original M2-brane spacetime (\ref{m2brane}).  It turns out that the supersymmetry condition (\ref{supercovariant}) can also be satisfied by taking the Kahler metric $g_{m\bar n}$ in (\ref{kahler}) to be flat, so that the relations (\ref{susy}) between $g$ and $H$ no longer hold.  The original M2-brane spacetimes are recovered in this way for $N=1$.  For $N>1$ one recovers the intersecting M2-brane spacetimes of \cite{Papadopoulos:1996uq} in this way.}.  The supercovariantly constant spinors are given by
\be
\epsilon=(E/f)^{-({N-1\over 6}+{a\over 4})}(\bar E/\bar f)^{-({N-1\over 6}-{a\over 4})}\epsilon_0,
\ee
where $\epsilon_0$ is a constant spinor satisfying the projection conditions (\ref{projection}).  
The source free equations of motion again reduce to equation (\ref{gaugeeom}).

The effective spatial metric and gauge potential for test M2-branes embedded in these spacetimes again have the form given in equation (\ref{effective}).  This implies that the warped Kahler forms $\phi$ in equation (\ref{warpedform}) are again generalized calibrating forms for test M2-branes.

\subskip{\bf New M5-brane Spacetimes:}
A more nontrivial application of our strategy is to start with worldvolume solitons calibrated by the square of the Kahler form $\phi={1\over 2}\omega\wedge\omega$.  Since this requires that the spatial dimension of the brane be at least $4$, in the context of M-theory we will be looking at M5-branes.  These spacetimes again will have two physical settings.  One could start with smoothed intersections of M5-branes that share a common string \cite{Gibbons:1999hm,Gauntlett:1999vk} in otherwise empty space  as in the discussion above equation (\ref{M5-branes}).  Alternatively, one can start with M5-branes wrapping a $(2,2)$ cycle of a Calabi-Yau manifold, leaving a string in the remaining noncompact directions.  

We build an FS ansatz similar to (\ref{kahler}) that reflects these new physical settings.  In particular, the calibrating form $\phi={1\over 2}\omega\wedge\omega$ of the worldvolume soliton is built into the $6$-form gauge potential.  Consider the $1/8$ supersymmetric case, corresponding to a $3$ complex dimensional space.  Accordingly, let
\be\label{squared}
\ba{l}
ds^2=H^{-2A}\left(-dt^2+dy^2\right )+2H^{-2B}g_{m\bar n}dz^mdz^{\bar n} + H^{2C}(\delta_{\alpha\beta}dx^\alpha dx^\beta)\\
A_{tym\bar nr\bar s}= c H^{-1}\left(g_{m\bar n}g_{r\bar s}-g_{m\bar s}g_{r\bar n}\right),\qquad c=\pm 1\\
\ea
\ee
where $m,n=1,\dots 3$ and $\alpha,\beta=1,2,3$.   We find that supersymmetry $(a)$ requires the projection conditions 
\be\label{newprojections}
\ba{cc}
\Gamma^{m\bar n}\epsilon= aH^{2B} g^{m\bar n}\epsilon, & a=\pm 1 \\
\Gamma^{ty}\epsilon= bH^{2A} \epsilon, & b=\pm 1 \\
\ea
\ee
with $bc=-1$; (b) fixes the values of the exponents to be $A=B=1/6$ and $C=1/3$; and (c) imposes the general relation (\ref{realrelation}) between $H$ and $g_{m\bar n}$.  The source free gauge field equations of motion again reduce to equation (\ref{gaugeeom}). We find that the supercovariantly constant spinors are given by
\be\label{spinors}
\epsilon=(E/f)^{-({1\over 12}+{a\over 4})}(\bar E/\bar f)^{-({1\over 12}-{a\over 4})}\epsilon_0,
\ee
where $\epsilon_0$ is a constant spinor satisfying the projection conditions (\ref{newprojections}).

Following \cite{Gutowski:1999tu}, we introduce a rescaled effective spatial metric $d\hat s^2$ for test M5-branes that are both static and translationally invariant in the $y$ direction.  The appropriate rescaling is 
$\hat G_{kl}= (-G_{tt}G_{yy})^{1/4} G_{kl}$, where $k,l$ run over all directions except $t,y$.  The $1/4$ power arises because these factors are now shared by the remaining $4$ spatial dimensions of the brane.  This yields
\be
d\hat s^2 = 2H^{-1/2}g_{m\bar n}dz^mdz^{\bar n}+ H^{1/2}(dx_8^2+dx_9^2+dx_{10}^2).
\ee
The calibrating form is then given by the expression $\phi={1\over 2}\omega\wedge\omega$ in terms of the Hermitian form
\be
\omega = cH^{-1/2}\omega_{\cal M}+H^{+1/2}dx_8\wedge dx_9.
\ee
The resulting form 
\be 
\phi = {1\over 2}H^{-1}\omega_{\cal M}\wedge\omega_{\cal M} +c\omega_{\cal M}\wedge dx_8 \wedge dx_9
\ee
is gauge equivalent to the effective spatial gauge potential $\hat A_{ijkl}=A_{tyijkl}$.  This can be seen by using the closure property of the Kahler metric $g_{m\bar n}$.

\section{Conclusion}
We have conjectured that the spacetime fields of $p$-brane worldvolume solitons are spacetimes of the FS type.  We have seen that thinking of FS spacetimes in terms of calibrations is useful both in understanding their structure and in generating new examples.  In this paper we have focused on Kahler calibrations.  As discussed above, we plan to investigate further examples in future work.

\vskip 0.2in\noindent
{\large\bf Acknowledgements:} We thank Robert Bryant and Don Marolf for helpful conversations.

%%%%%%%%%%%%%%%%%%%%

\end{document}